\documentclass[aps,prb,twocolumn,superscriptaddress]{revtex4-1}
\usepackage{fancyhdr}
\usepackage{color}
\usepackage[final]{graphicx}
\usepackage{graphicx}
\usepackage{sidecap}
\usepackage{wrapfig}
\usepackage[usenames,dvipsnames]{xcolor}
\usepackage{palatino}
\usepackage{hyperref}
\usepackage{multirow}
\usepackage{caption}
\usepackage{array}
\usepackage{geometry}
\usepackage{mathtools}
\usepackage{fancyhdr}
\usepackage{braket}
\usepackage{cprotect}
\usepackage{eurosym}
\usepackage{tabularx}
\usepackage{empheq}
\usepackage{amsmath}
\usepackage{color, colortbl}
\usepackage[para]{footmisc}
\usepackage[capitalize]{cleveref}
\usepackage{multirow}
\usepackage{footnote}
\usepackage[normalem]{ulem}

\begin{document}


\title{Density functional perturbation theory within non-collinear magnetism}


\author{Fabio Ricci}
\affiliation{CESAM-QMAT Physique Th\'eorique des Mat\'eriaux, Universit\'e de Li\`ege, B-4000 Sart-Tilman, Belgium}
\email{corresponding author: fabio.ricci@uliege.be}
\author{Sergei Prokhorenko}
\affiliation{CESAM-QMAT Physique Th\'eorique des Mat\'eriaux, Universit\'e de Li\`ege, B-4000 Sart-Tilman, Belgium}
\author{Marc Torrent}
\affiliation{CEA, DAM, DIF, F-91297 Arpajon, France}
\author{Matthieu J. Verstraete}
\affiliation{CESAM-QMAT-nanomat, and European Theoretical Spectroscopy Facility Universit\'e de Li\`ege, B-4000 Sart-Tilman, Belgium}
\author{Eric Bousquet}
\affiliation{CESAM-QMAT Physique Th\'eorique des Mat\'eriaux, Universit\'e de Li\`ege, B-4000 Sart-Tilman, Belgium}
\email{corresponding author: eric.bousquet@uliege.be}


\date{\today}

\begin{abstract}
We extend the density functional perturbation theory formalism to the case of non-collinear magnetism. 
The main problem comes with the exchange-correlation (XC) potential derivatives, which are the only ones that are affected by the non-collinearity of the system. 
Most of the present XC functionals are constructed at the collinear level, such that the off-diagonal (containing magnetization densities along x and y directions) derivatives cannot be calculated simply in the non-collinear framework. 
To solve this problem, we consider here possibilities to transform the non-collinear XC derivatives to a local collinear basis, where the $z$ axis is aligned with the local magnetization at each point.
The two methods we explore are i) expanding the spin rotation matrix as a Taylor series, ii) evaluating explicitly the XC for the local density approximation through an analytical expression of the expansion terms. 
We compare the two methods and describe their practical implementation.
We show their application for atomic displacement and electric field perturbations at the second order, within the norm-conserving pseudopotential methods.
\end{abstract}


\maketitle

\section{Introduction}

The density functional perturbation theory (DFPT) is the application of linear response theory to the density functional theory (DFT) formalism.
Its development traces back to the birth of many DFT codes during the 80's \cite{baroni2001,Gonze1995b,baroni1987}.
The first applications were for the calculations of phonon spectra, dielectric constants, Born effective charges and related properties \cite{Gonze1997}.
These responses require to expand the DFT as a function of atomic displacements and electric field perturbations.
Later on, strain perturbations were added to the list, which allows the calculation of elastic and piezoelectric tensors \cite{hamann2005,wu2005}.
DFPT represented a breakthrough in the calculation of many physical properties since, once the equations are implemented, the resulting work-flow is very simple for the end user; this also simplifies high-throughput calculations \cite{petousis2016,hautier2012}. 

However, in spite of the past decade's increase of interest for non-collinear magnetic materials, DFPT has not yet been generalized for non-collinear magnets. 
This is due to the fact that non-collinear magnetism within DFT is much more computationally demanding (the density has four components instead of two for collinear magnetism) and it requires to include the spin-orbit coupling (SOC), which, in present implementations of DFT codes, often reduces the number of usable symmetries \cite{peralta2007,hobbs2000,kubler1988}. 
Additionally, non-collinear magnetism with SOC works in very low energy ranges (from 1 to hundreds of $\mu$eV), which require highly accurate simulations \cite{bousquet2010}. Non-collinear ground state (GS) calculations (spin and lattice) have been first developed in the 80's \cite{kubler1988} but became widespread only recently thanks to increased supercomputer power and the improvement of code scalability \cite{naveh2009,peralta2007,hobbs2000, verstraete_2008_so_phonons, dalcorso2010}.
It is thus timely to extend DFPT to non-collinear magnetism, for phonon spectra, elasticity or dielectric responses.

In this paper, we derive the formalism of DFPT in the presence of non-collinear magnetism. The main difficulty is represented by the XC term, for which the functionals have been developed using a collinear framework. Only a few attempts have been made to go beyond this and parameterize directly a non-collinear functional~\cite{eich_2013_noncollinear_spin_xc, eich_2013_noncollinear_spin_xc2, bulik2013, scalmani2012}.
We show here two different approaches to perform the energy derivatives in the non-collinear regime starting with the collinear $E_{xc}$ functionals.
We start with a rapid reminder of collinear and of non-collinear DFT formalisms, followed by the description of two formally equivalent methods to derive the XC potential from collinear $E_{xc}$ functionals. 
We also show the specific derivations (using norm-conserving pseudo potentials), for atomic displacement and electric field perturbations. 
In each case, we show how the two methods compare in precision and efficiency.

\section{Basics of DFPT}
\label{sec:DFPTbase}

The framework of DFT\cite{Kohn1965, hohenberg_1964} allows to solve a many body problem as a single particle in an effective self-consistent potential. In this formalism \cite{Kohn1999}, the reference Ground State (GS) energy is estimated by minimizing with respect to the Kohn-Sham (KS) wavefunctions $\psi_{\alpha}$ the following expression:
\begin{equation}\label{eq:GSene}
{\rm E}[\rho]=\sum_n\braket{\psi_{n}|T+V_{ext}|\psi_{n}}+{\rm E}_{\rm H}[\rho] + {\rm E}_{xc}[\rho]  .
\end{equation}
Here, $n$ runs over all the occupied states, $T$ is the kinetic energy operator, $v_{ext}$ is the external potential (the sum of ionic potentials, which can be non-local, and eventual applied fields), $E_{\rm H}$ and $E_{xc}$ are the interaction energy terms, namely the Hartree ($H$) and XC functionals. These two last terms contain the approximations to the many body electronic interactions, as functionals of the density operator:
\begin{equation}
\hat{\rho}({\bf r})=\sum_{n}\psi_{n}^{*}({\bf r})\psi_{n}({\bf r}).
\end{equation}
The Hamiltonian related to Eq.~(\ref{eq:GSene}), needed to self-consistently solve the KS equation, is:
\begin{equation}\label{eq:hamilt}
{\rm H}= T+V_{ext}+V_{H}+V_{xc}\ .
\end{equation}
The solution of this equation depends on the way the last term is approximated in the so-called XC potential $V_{xc}$. Interestingly, $V_{xc}$ can be written as the functional derivative of the XC energy with respect to the density:
\begin{equation}\label{eq:xcpotcol}
V_{xc}=\dfrac{\delta {\rm E}_{xc}[\rho]}{\delta \rho({\bf r})} \quad .
\end{equation}

When the system is subject to a perturbation, its effects can be related to a (small) parameter $\lambda$ on which any generic observable $X(\lambda)$ depends. Expanding explicitly the perturbation series we have:
\begin{equation}
X(\lambda)=X^{(0)}+\lambda X^{(1)}+\lambda^2 X^{(2)}+\lambda^3 X^{(3)}+...
\end{equation}
where the expansion coefficients are related to the $N^{\rm th}$ order derivative of $X(\lambda)$ with respect to $\lambda$ by a scalar coefficient:
\begin{equation}
X^{(N)}=\dfrac{1}{N!}\dfrac{d^NX}{d\lambda^N}\Bigr|_{\lambda=0} \quad .
\end{equation}
Our objective is to evaluate the modification of the GS energy E or density $\rho({\bf r})$ caused by the presence of the external perturbation, through the computation of this expansion.

At the $i^{\rm th}$ generic order in $\lambda$, the XC potential can be written as:
\begin{equation}
V_{xc}^{(i)}=\left(\dfrac{\delta E_{xc}[\rho(\lambda)]}{\delta \rho({\bf r})}\right)^{(i)}=\dfrac{1}{i!}\left(\dfrac{d^{(i)}}{d\lambda^{(i)}}\dfrac{\delta E_{xc}[\rho(\lambda)]}{{\delta \rho({\bf r})}}\right)\Bigr|_{\lambda=0}\ .
\end{equation}
Truncating the expansion at the first order, the equation that has to be self-consistently solved is obtained expanding the single particle KS equation~\cite{Gonze1995b,Gonze1997}:
\begin{equation}\label{KSeq}
\left[{\rm H}(\lambda)-E_{n}(\lambda)\right]\ket{\psi_{n}(\lambda)}=0\ ,
\end{equation}
giving the so-called Sternheimer equation \cite{Sternheimer1954}:
\begin{equation}
P_c(H-E_{n})^{(0)}P_c\ket{\psi^{(1)}_{n}}+P_c(H-E_{n})^{(1)}\ket{\psi^{(0)}_{n}}=0\  ;
\end{equation}
here, $P_c$ is the projector upon the empty states (conduction bands), $H^{(0)}$, $E_{\alpha}^{(0)}$ and $\psi^{(0)}$ are, respectively, the GS Hamiltonian, eigenvalues and wavefunctions and $H^{(1)}$ is the first-order Hamiltonian:
\begin{eqnarray}
{\rm H}^{(1)}&=&V_{ext}^{(1)}+V_{xc}^{(1)}=V_{ext}^{(1)}+\nonumber\\
&+&\int\dfrac{\delta^2E_{xc}}{\delta\rho({\bf r})\ \delta\rho({\bf r}')}\Bigr|_{\rho^{(0)}}\rho^{(1)}({\bf r}')\ d{\bf r}'+\nonumber\\
&+&\dfrac{d}{d\lambda}\ \dfrac{\delta E_{xc}}{\delta \rho({\bf r})}\Bigr|_{\rho^{(0)}}\ .
\end{eqnarray}
The orthonormalization condition that holds for the GS wavefunctions $\braket{\psi_{n}|\psi_{n}}=1$, becomes to first order:
\begin{equation}
\braket{\psi^{(0)}_{n}|\psi^{(1)}_{n}}+\braket{\psi^{(1)}_{n}|\psi^{(0)}_{n}}=0  .
\end{equation} 

Once the framework has been specified (DFT for the search of the GS and DFPT to explore the effects of the perturbations), it is very useful to write the operators in the spinorial form: each operator belonging to the KS equation is represented as a 2$\times$2 matrix. In this way, we can rewrite the Eq. (\ref{KSeq}) as:
\begin{equation}
\left( \begin{array}{cc}
H^{\alpha\alpha} & V_{xc}^{\alpha\beta} \\[15px]
V_{xc}^{\beta\alpha} & H^{\beta\beta}  \end{array} \right)\ \left( \begin{array}{c} \ket{\psi_n^\alpha}\\[16px] \ket{\psi_n^{\beta}}\end{array} \right)\ = E_n\left( \begin{array}{c} \ket{\psi_n^\alpha}\\[16px] \ket{\psi_n^{\beta}}\end{array} \right)  ,
\end{equation}
here, the indices $\alpha$ and $\beta$ run over the up and down spin components (or major and minor Pauli spinors) and $n$ over the bands. As we can see, the off-diagonal part of the hamiltonian comes only from the XC potential: all other terms of H (kinetic energy, Hartree and pseudo-potential components) do not cross-couple up and down spin channels. At the zeroth order, the general $\alpha\beta$ element of the XC potential is obtained simply re-writing Eq. (\ref{eq:xcpotcol}):
\begin{equation}
V_{xc}^{\alpha\beta}\left[\hat{\rho}({\bf r})\right]=\dfrac{\delta E_{xc}\left[\hat{\rho}({\bf r})\right]}{\delta \rho^{\alpha\beta}({\bf r})}  .
\end{equation}
and since the (spin) density matrix is diagonal ($\rho_{\alpha\beta}({\bf r})=\hat{\rho}^{\alpha\beta}\delta_{\alpha\beta}$) only if the magnetization is directed along the quantization axis $z$ (then the magnetization ${\bf m}({\bf r})$ = m$_z({\bf r})$), it is useful to diagonalize the 2$\times$2 matrix in the previous section, finding the transformation $\hat{U}$ (often referred to as the ``spin-$1/2$ rotation matrix''). 
Using the new basis, the XC potential becomes:
\begin{eqnarray}
V_{xc}^{\alpha\beta}&=&\dfrac{1}{2}\left[\dfrac{\delta E_{xc}}{\delta \rho^{\uparrow}}+\dfrac{\delta E_{xc}}{\delta \rho^{\downarrow}}\right]\delta_{\alpha\beta}+ \nonumber\\
&+&\dfrac{1}{2}\left[\dfrac{\delta E_{xc}}{\delta \rho^{\uparrow}}-\dfrac{\delta E_{xc}}{\delta \rho^{\downarrow}}\right]\left(\hat{U}^{\dagger}\boldsymbol{\sigma}_z\hat{U}\right)_{\alpha\beta}=\\
&=&\dfrac{1}{2}\left[\dfrac{\delta E_{xc}}{\delta \rho^{\uparrow}}+\dfrac{\delta E_{xc}}{\delta \rho^{\downarrow}}\right]\delta_{\alpha\beta}+ \nonumber\\
&+&\dfrac{1}{2}\left[\dfrac{\delta E_{xc}}{\delta \rho^{\uparrow}}-\dfrac{\delta E_{xc}}{\delta \rho^{\downarrow}}\right]\hat{{\bf m}}\cdot\boldsymbol{\sigma}_{\alpha\beta}  .\label{eq:pot0}
\end{eqnarray}
where $\rho^{\uparrow(\downarrow)}({\bf r})=1/2\left(n\pm m\right)$, where $n$ is the charge density; $\delta_{\alpha\beta}$ is the Kronecker delta, $\hat{{\bf m}}={\bf m}/m$ is the magnetization versor and $m$ the norm of ${\bf m}$; $\boldsymbol{\sigma}_{\alpha\beta}(\sigma^x_{\alpha\beta},\sigma^y_{\alpha\beta},\sigma^z_{\alpha\beta})$ are the Pauli matrices. 

In the following, we extend the DFPT formalism to non-collinear magnetism using the spinorial wavefunctions. 
With respect to collinear magnetism, {\em i. e.}, where the direction of all the spins is along a well defined quantization axis, there are at least two issues to be addressed: 1) the calculation of the derivative of the XC potential $V^{(1)}_{xc\ \alpha\beta}$ and 2) the computation of the density derivatives $\rho^{(1)}_{\alpha\beta}$ from the derivative of the KS wavefunctions $\psi^{(1)}_{n\ \alpha}$.
 
 \section{Non-collinear DFT framework}

Our goal is now to generalize the DFPT equations with spinorial wavefunctions, to take into account the NCM and thus a generic magnetization direction.
The main concern of such a generalization is to treat the derivatives of the XC term (see above).
A first problem arises from the fact that in spin-dependent DFT the XC functionals are constructed on a collinear basis where the magnetization is considered to be have the same orientation at each point in space.
The standard way to handle non-collinear DFT with collinear XC functionals is to locally rotate the magnetization density with respect the quantization axis at each point in space, so that the derivatives of the XC term can be treated as locally collinear~\cite{hobbs2000}, and then rotated back to the ``laboratory frame''.
To avoid this ``local collinearity approximation'', some attempts have been made to build specific non-collinear XC functionals~\cite{eich_2013_noncollinear_spin_xc, eich_2013_noncollinear_spin_xc2, bulik2013,scalmani2012}.
In this paper, we will restrict ourselves to the case of  ``local collinearity approximation'' and use the common collinear XC functionals.

In a collinear magnetic system, the magnetization at each point {\bf r} in space is considered to be directed along a fixed quantization axis (usually $z$). 
In this picture, the density matrix has two components only, which are often specified using $\alpha,\beta=1,2$, to indicate the spin up $\ket{1}$ and down $\ket{2}$ states:
\begin{eqnarray}
\rho_{11}&=&\rho^{\uparrow}=\dfrac{1}{2}\left(n+m_z\right)\label{rhoupup}\\
\rho_{22}&=&\rho^{\downarrow}=\dfrac{1}{2}\left(n-m_z\right)\label{rhodndn}\\
\rho_{12}&=&\rho_{21}=0.\nonumber
\end{eqnarray}
In these expressions and the following ones, we omit the $^{(0)}$ or $^{(1)}$ to indicate the order of the perturbation, since the form of the density matrix is generally valid. 

In the case of NCM, the density matrix operator can be written in terms of the GS spinorial wavefunctions $\psi^{(0)}$ as follows:
\begin{equation}\label{eq:rho0}
\hat{\rho}=\ket{\psi^{(0)}}\bra{\psi^{(0)}} .
\end{equation}
In the representation of the spin, projecting the spinors on the orthonormal spin basis $\ket{\alpha}$ and $\ket{\beta}$, $\hat{\rho}_{\alpha\beta}$ can be evaluated explicitly in terms of spin components:
\begin{equation}
\rho_{\alpha\beta}=\braket{\alpha|\psi^{(0)}}\braket{\psi^{(0)}|\beta} ,
\end{equation}
or in matrix form:
\begin{eqnarray}\label{eq:rhonew}
\rho&=&
\left( \begin{array}{cc}
\psi^{(0)}_1\psi_1^{(0)*} & \psi^{(0)}_1\psi_2^{(0)*} \\[1em]
\psi^{(0)}_2\psi_1^{(0)*} & \psi_2^{(0)}\psi_2^{(0)*} \end{array} \right)  \\
&=&\dfrac{1}{2}\left( \begin{array}{cc}\label{eq:ropaul}
n + m_z & m_x-i\ m_y \\[1em]
m_x + i\ m_y & n-m_z \end{array} \right)\ ;
\end{eqnarray}
where $\braket{\alpha|\psi^{(0)}}=\psi_{\alpha}^{(0)}$ and $\braket{\psi^{(0)}|\beta}=\psi_{\beta}^{(0)*}$, $m_i$ is the $i=x,y,z$ component of the spin vector, or magnetization density, and $n$ is the electronic density.
This expression for the density matrix is defined by its relation with the Pauli matrices $\boldsymbol{\sigma}_{\alpha\beta}$ \cite{Ziegler2006}, which gives the following expression:
\begin{eqnarray}
\rho_{\alpha\beta}&=&\dfrac{1}{2}\left[n\ \delta_{\alpha\beta}+{\bf m}\cdot{\boldsymbol\sigma}_{\alpha\beta}\right]=\\
&=&\dfrac{1}{2} \left[n\ \delta_{\alpha\beta}+\left(\sum_{i=x,y,z}m_i\ \sigma^{i}_{\alpha\beta}\right)\right] .
\end{eqnarray}
where the electronic and magnetization components can be written using Eq. (\ref{eq:rhonew}):
\begin{eqnarray}
n&=&\left(\psi_1^{(0)*}\psi_1^{(0)}+\psi_2^{(0)*}\psi_2^{(0)}\right) \label{eq:density0}\\
m_x&=&\left(\psi_1^{(0)*}\psi_2^{(0)}+\psi_2^{(0)*}\psi_1^{(0)}\right)\\
m_y&=&i\left(\psi_2^{(0)*}\psi_1^{(0)}-\psi_1^{(0)*}\psi_2^{(0)}\right)\\
m_z&=&\left(\psi_1^{(0)*}\psi_1^{(0)}-\psi_2^{(0)*}\psi_2^{(0)}\right)
\end{eqnarray}

The expression of the GS charge density $n$ in Eq. (\ref{eq:density0}) can be written as specified in Ref. \cite{Gonze1995b}:
\begin{equation}
n^{(0)}({\bf r})=\dfrac{1}{(2\pi)^3}\int_{\textrm{BZ}}\sum_i^{occ}\sum_{s=1}^2\ \psi_{i{\bf k}s}^{(0)*}({\bf r})\ \psi_{i{\bf k}s}^{(0)}({\bf r})\ d{\bf k},
\end{equation}
where $i$ runs over all the occupied states, $s$ is the spin component (if the system is non-magnetic $s=1$) and the integral is performed on the whole Brillouin Zone (BZ). 

\section{DFPT within non-collinear magnetism}

In order to address a NCM problem, we want to map the overall 2$\times$2 density matrix form (Eq. (\ref{eq:ropaul})) onto a local collinear case.
In the following we develop two possibilities to do this mapping. Both techniques rely on the idea of first evaluating the perturbed XC potential ($V^{(1)}_{xc}({\bf r})$) in a local reference frame, whose $z$-axis is aligned with the direction of \emph{local perturbed magnetization} at each point {\bf r} of the real space. 
Such a choice of local coordinates reduces the problem to that of collinear magnetism, and after the evaluation of the corresponding ``collinear'' $V^{(1)}_{xc}({\bf r})$ values, we can restore its non-collinear form applying the backwards rotation that aligns the local reference frame with the globally defined Cartesian axes. 
The first order potential can be then evaluated based on either ({\bf i}) the $\lambda$ expansion of the rotation matrices and corresponding series of the collinear $V^{(1)}_{xc}$ expression or ({\bf ii}) by direct evaluation of the $\lambda$ series of the final non-collinear $V^{(1)}_{xc}$ expression.  Essentially, we develop either the expansion of the general expression (case ({\bf i})) or of the final result (case ({\bf ii})). 
Although both methods give the same results, the first method ({\bf i}) is more general: the procedure to evaluate $V^{(1)}_{xc}$ can be applied without any modification of the XC potential (even a non-diagonal one), while the second method requires the explicit re-evaluation of the $\lambda$ series for each particular choice of XC flavour (LDA or GGA). For the sake of completeness we provide below the development of both schemes (see sections \ref{sec:methodROT} and \ref{sec:methodBfield} for methods ({\bf i}) (and an alternative one that we call ({\bf i}')) and ({\bf ii}), respectively). 
Furthermore, implementation of both methods allows us to validate the correctness of the implemented code by contrasting results obtained using the two different techniques.

\subsection{(i) Expansion of the rotation matrix method}
\label{sec:methodROT}
\subsubsection{Magnetization density local transformation}
To project the non-collinear density onto the local axis of magnetization, we need to determine the direction of the local spin-quantization axis to build a diagonal collinear density.
In the case of a general GS density matrix $\hat{\rho}$ of the form Eq. (\ref{eq:ropaul}), the eigenvalues are given by $\rho_{\uparrow,\downarrow}=1/2\left(n\pm|m|\right)$ and the two corresponding normalized and complex eigenvectors are:
\begin{eqnarray}
N_1&=&\dfrac{1}{D_1}\binom{m_z+m}{m_x+i\ m_y}\\
N_2&=&\dfrac{1}{D_2}\binom{m_z-m}{m_x+i\ m_y}\ , \ {\rm with}\\
D_1&=&\sqrt{\left|m+m_z\right|^2+\left|m_x+i\ m_y\right|^2}\\
D_2&=&\sqrt{\left|-m+m_z\right|^2+\left|m_x+i\ m_y\right|^2}
\end{eqnarray}
where $m = |{\bf m}| = \sqrt{m_x^2 + m_y^2 + m_z^2}$.
The two $N_1$ and $N_2$ eigenvectors define the spin-1/2 unitary transformation matrix $\hat{U}$ diagonalizing $\hat{\rho}$. Then, the transformation can be detailed as:
\begin{equation}\label{eq:diag0}
\sum_{\alpha\beta}U_{i\alpha}^{\dagger}\ \hat{\rho}^{\alpha\beta}\ U_{\beta j}=\rho_{ij}\ \delta_{ij} ,
\end{equation}
here $\alpha,\beta,i,j$ are indices running on the two possible spin states $\ket{1}$ or $\ket{2}$ (as previously defined).
The expression of $\hat{U}$ is thus:
\begin{equation}
\hat{U}=\left( \begin{array}{cc}
\dfrac{m_z + m}{D_1} & \dfrac{m_z-m}{D_2} \\[15px]
\dfrac{m_x + i\ m_y}{D_1} & \dfrac{m_x + i\ m_y}{D_2} \end{array} \right).
\end{equation}
The inverse of this matrix, such that $\hat{U}^{-1}\hat{U}={\bf I}$, with ${\bf I}$ the identity matrix, is:
\begin{equation}
\hat{U}^{-1}=\left( \begin{array}{cc}
\dfrac{D_1}{2m} & \dfrac{\left(m-m_z\right)D_1}{2m\left(m_x+i\ m_y\right)} \\[15px]
-\dfrac{D_2}{2m} & \dfrac{\left(m+m_z\right)D_2}{2m\left(m_x+i\ m_y\right)} \end{array} \right).
\end{equation}
Moreover, at all orders in $\lambda$, we have the unitarity property: $\hat{U}^{\dagger}=\hat{U}^{-1}$.
Importantly, in the last two expressions the presence of the ${\bf m}$ components in the denominator does not cause a divergence if ${\bf m}=0$, since the dependence on the magnetization of the numerator is of the same order. To the first order in the perturbation, the unitarity condition continues to hold:
\begin{equation}
\sum_{\alpha}U_{i\alpha}U_{\alpha j}^{\dagger}=\sum_{\alpha}U^{\dagger}_{i\alpha}U_{\alpha j}=\delta_{ij} \ , 
\end{equation}
and can be expanded:
\begin{equation}
\resizebox{.85 \linewidth}{!}{
$\sum_{\alpha}\left(U_{i\alpha}^{(0)}+\lambda U_{i\alpha}^{(1)}\right)^{\dagger}\left(U_{i\alpha}^{(0)}+\lambda U_{i\alpha}^{(1)}\right)=\delta_{ij}$  .
}
\end{equation}
Developing this expression, neglecting the second order perturbations ($\lambda^2$) and higher terms, we obtain the following relationship:
\begin{equation}\label{eq:Uunit1}
\sum_{\alpha}U_{i\alpha}^{\dagger(1)}U_{\alpha j}^{(0)}=-\sum_{\alpha}U_{i\alpha}^{\dagger(0)}U_{\alpha j}^{(1)} .
\end{equation}
All these properties will be exploited in order to obtain the expression of the perturbed XC potential and density matrix in the following part of the manuscript.

\subsubsection{XC potential and density matrix}
\label{sec:methodBfield}

As we have seen for a general observable $X$ in Sec. \ref{sec:DFPTbase}, the perturbed XC potential (in the following we will omit the ``$xc$'' subscript in  $V_{xc}$) can be expanded as a Taylor series in $\lambda$ and truncated at the first order:
\begin{equation}\label{eq:vfirst1}
{V}_{\alpha\beta}=V^{(0)}_{\alpha\beta}+\lambda\ V^{(1)}_{\alpha\beta} +o(\lambda^2) .
\end{equation}
At each point in space the equation to locally diagonalize the total XC potential is:
\begin{equation}\label{eq:vdiag1}
\sum_{\alpha\beta}{U}_{i\alpha}^{\dagger}\ {V}_{\alpha\beta}\ {U}_{\beta j}={V}_{i}\ \delta_{ij}  .
\end{equation}
Here, ${U}$ is a general transformation matrix to be determined.
Substituting now the expression in Eq. (\ref{eq:vfirst1}), truncated at the first order, into the last equation, we obtain the complete expression needed to rotate the reference frame along the local direction where the XC potential is diagonal:
\begin{eqnarray}
\sum_{\alpha\beta}\left(U_{i\alpha}^{(0)}+\lambda U_{i\alpha}^{(1)}\right)^{\dagger}\left(V^{(0)}_{\alpha\beta}+\lambda V^{(1)}_{\alpha\beta}\right)&&\left(U_{i\alpha}^{(0)}+\lambda U_{i\alpha}^{(1)}\right)= \nonumber \\
=\left(V^{(0)}_{ij}+\lambda V^{(1)}_{ij}\right)\delta_{ij} &&
\end{eqnarray}
In passing we note that, for general non-collinear XC functionals, the relation between diagonalizing $\rho$ and $V$ might be more complex.
It is interesting to separate the terms belonging to the last expression, as a function of their order in $\lambda$:
\begin{eqnarray}
&\xRightarrow{0^{th}}&\ \sum_{\alpha\beta}U_{i\alpha}^{\dagger(0)}V_{\alpha\beta}^{(0)}U_{\beta j}^{(0)}-V_i^{(0)}\delta_{ij}\ +\label{eq:v0order}\\
&\xRightarrow{1^{st}}&\ +\ \lambda\sum_{\alpha\beta}\left[U_{i\alpha}^{\dagger(1)}V_{\alpha\beta}^{(0)}U_{\beta j}^{(0)}+U_{i\alpha}^{\dagger(0)}V_{\alpha\beta}^{(0)}U_{\beta j}^{(1)}\right.+ \nonumber\\
&+&\left.U_{i\alpha}^{\dagger(0)}V_{\alpha\beta}^{(1)}U_{\beta j}^{(0)}-V_i^{(1)}\delta_{ij}\right]+\label{eq:v1order}\\
&\xRightarrow{\ge 2^{nd}}&\ +\ o(\lambda^2) = 0  .
\end{eqnarray}
From Eq. (\ref{eq:v0order}) we can extract two useful properties of the zeroth order transformation equation:
\begin{eqnarray}
&\textrm{{\bf a.}}&  \sum_{\beta}V_{\alpha\beta}^{(0)}U_{\beta j}^{(0)}=U_{\alpha j}^{(0)}V_{j}^{(0)} \label{eq:vprop2a}\\
&\textrm{{\bf b.}}&   \sum_{\alpha}U_{i\alpha}^{\dagger(0)}V_{\alpha\beta}^{(0)}=V_i^{(0)}U_{i\beta}^{\dagger(0)} \label{eq:vprop2b} .
\end{eqnarray}
We can rewrite the Eq. (\ref{eq:v1order}) using the unitary properties (\ref{eq:Uunit1}) and the previous relations (\ref{eq:vprop2a}) and (\ref{eq:vprop2b}) as follows:
\begin{eqnarray}
&&\sum_{\alpha}U_{i\alpha}^{\dagger(1)}U_{\alpha j}^{(0)}V_j^{(0)}+\sum_{\beta}V_i^{(0)}U_{i\beta}^{\dagger(0)}U_{\beta j}^{(1)} + \nonumber \\
&+&\sum_{\alpha\beta}U_{i\alpha}^{\dagger(0)}V_{\alpha\beta}^{(1)}U_{\beta j}^{(0)}=V_i^{(1)}\delta_{ij} ,
\end{eqnarray}
and after some algebraic manipulation we get:
\begin{eqnarray}\label{eq:veqsol}
\sum_{\alpha}U_{i\alpha}^{\dagger(1)}U_{\alpha j}^{(0)}\left[V_j^{(0)}-V_i^{(0)}\right]&+&\nonumber\\
+\ \sum_{\alpha\beta}U_{i\alpha}^{\dagger(0)}V_{\alpha\beta}^{(1)}U_{\beta j}^{(0)}&=&V_i^{(1)}\delta_{ij}\ .
\end{eqnarray}

The last equation can be represented symbolically:
\begin{equation}
\left( \begin{array}{cc} 0 & \bigtriangleup \\[8px]  \bigtriangleup^{*} & 0  \end{array} \right)+\left( \begin{array}{cc} \odot & - \bigtriangleup \\[8px] - \bigtriangleup^{*} & \odot \end{array} \right) =
\left( \begin{array}{cc} \odot & 0 \\[8px] 0 & \odot  \end{array} \right)
\end{equation}
As we can see, the first term on the left-hand side is a purely off-diagonal matrix which is needed to compensate the off-diagonal terms present in the second term to guarantee that the right-hand side term is diagonal. Interestingly, this implies that the first order transformation matrix $\hat{U}^{(1)}$ can never be neglected since its presence ensures the unitarity of the transformation. Determining the elements of $\hat{U}^{(1)}$ requires some algebraic effort, which can be circumvented by exploiting the analogous of Eq. (\ref{eq:veqsol}) for the density matrix $\hat{\rho}$:
\begin{eqnarray}\label{eq:eqsol}
\sum_{\alpha}U_{i\alpha}^{\dagger(1)}U_{\alpha j}^{(0)}\left[\rho_j^{(0)}-\rho_i^{(0)}\right]&+&\nonumber\\
+\ \sum_{\alpha\beta}U_{i\alpha}^{\dagger(0)}\rho_{\alpha\beta}^{(1)}U_{\beta j}^{(0)}&=&\rho_i^{(1)}\delta_{ij}
\end{eqnarray}
In fact, the same mathematical process from Eq. (\ref{eq:vfirst1}) to (\ref{eq:veqsol}) is valid for the density operator as well. 
In this way, it holds for both Eq.s (\ref{eq:veqsol}) and (\ref{eq:eqsol}) that the quantity
\begin{equation}
M=-M^{\dagger}=\sum_{\alpha}U_{i\alpha}^{\dagger (1)}U_{\alpha j}^{(0)}
\end{equation}
can be written explicitly from the density matrix Eq. (\ref{eq:eqsol}) as:
\begin{equation}
M_{ij}=
\begin{cases}
i=j\quad\quad \quad\quad\quad0\\[2em]
i\neq j\quad\quad   -\dfrac{\tilde{\rho}^{(1)}_{ij}}{\rho_j^{(0)}-\rho_i^{(0)}}
\end{cases}
\end{equation}
where we have defined
\begin{equation}\label{eq:rotilde}
\tilde{\rho}^{(1)}_{ij}=\sum_{\alpha\beta}U_{i\alpha}^{\dagger(0)}\rho_{\alpha\beta}^{(1)}U_{\beta j}^{(0)}.
\end{equation}
The knowledge of this quantity fully solves the system of Eq.s (\ref{eq:veqsol}) and (\ref{eq:eqsol}), indeed, to obtain the transformed XC potential in the Cartesian system of reference we shift the second term from the left-hand to the right-hand side of Eq. (\ref{eq:veqsol}) and apply the inverse transformation.
Similarly to Eq. (\ref{eq:rotilde}), we can define an analogous expression for XC potential:
\begin{equation}
\tilde{V}^{(1)}_{ij}=\sum_{\alpha\beta}U_{i\alpha}^{\dagger(0)}V_{\alpha\beta}^{(1)}U_{\beta j}^{(0)}.
\end{equation}
Now, we can rewrite Eq. (\ref{eq:veqsol}) and find the expression for the perturbed potential in the local reference system:
\begin{equation}
\label{eq:v1tilde}
\tilde{V}^{(1)}_{ij}=V_i^{(1)}\delta_{ij}-M_{ij}\left(V_j^{(0)}-V_i^{(0)}\right)
\end{equation}
In this way, the final expression for the matrix elements is:
\begin{equation}
\tilde{V}_{ij}^{(1)}=\label{eq:velnor}
\begin{cases}
i=j\quad\quad\quad\quad\quad         V_i^{(1)}\\[2em]
i\neq j\quad\quad\quad   \dfrac{\tilde{\rho}^{(1)}_{ij}}{\rho_j^{(0)}-\rho_i^{(0)}}\left(V_j^{(0)}-V_i^{(0)}\right)
\end{cases}
\end{equation}
and its expression in the global Cartesian reference requires to apply the inverse transformation to Eq. (\ref{eq:velnor}):
\begin{equation}
\label{eq:resROT}
V_{\alpha\beta}^{(1)}=\sum_{ij}U_{\alpha i}^{(0)}\left[V_i^{(1)}\delta_{ij}-M_{ij}\left(V_j^{(0)}-V_i^{(0)}\right)\right]U_{j \beta}^{\dagger(0)}\ .
\end{equation}

This method and the expression for the perturbed XC potential is completely general and independent of its flavour (LSDA or GGA or others, for example).

It is possible, moreover, to find the analytical expressions of the $U^{(0)}$ and $U^{(1)}$ transformation matrices in terms of the Pauli matrices ${\boldsymbol\sigma_{\alpha\beta}}$, {\em i. e.}, the generators of the Lie group SU(2). Specifically, we can write:
\begin{equation}\label{eq:usu2a}
U=\exp{\left[-i\dfrac{\theta}{2}\left({\boldsymbol\sigma}\cdot {\hat n}\right)\right]}=\cos{\left(\dfrac{\theta}{2}\right)}-i\sin{\left(\dfrac{\theta}{2}\right)}\left({\boldsymbol\sigma}\cdot {\hat n}\right)
\end{equation}
where ${\hat n}$ denotes the rotation axis direction and $\theta$ the rotation angle (in the following, the $(0)$ superscript will be omitted for 0$^{th}$ order quantities). In particular, these two quantities can be written in terms of the magnetization components
\begin{equation}
\resizebox{\columnwidth}{!}{
${\hat n}=\dfrac{\hat{z}\times {\bf m}}{\left|\hat{z}\times {\bf m}\right|}=\left(-\dfrac{m_y}{\sqrt{m_x^2+m_y^2}};\ \dfrac{m_x}{\sqrt{m_x^2+m_y^2}};\ 0\right)$\label{eq:usu2b}
}
\end{equation}
\begin{flalign}
& \theta=\arccos{\dfrac{m_z}{m}}\label{eq:usu2c}
\end{flalign}
thus defining the rotation operation from the local spin-coordinate frame (with the local $z$ axis aligned within the local magnetization direction) to the global spin-coordinate reference system. The use of Eq.s (\ref{eq:usu2a}), (\ref{eq:usu2b}) and (\ref{eq:usu2c}) allows to obtain an analytical expression for the $U^{(1)}$ and $U^{\dagger(1)}$ matrices:
\begin{equation}
\begin{split}
U^{(1)} =& -\dfrac{\sin{(\theta/2)}+i\cos{(\theta/2)}({\boldsymbol\sigma}\cdot {\hat n})}{2}\theta^{(1)}+ \\
& -i\sin{(\theta/2)}({\boldsymbol\sigma}\cdot {\hat n}^{(1)})\label{eq:usu2d}
\end{split}
\end{equation}
\begin{equation}
\begin{split}
U^{\dagger(1)}=&-\dfrac{\sin{(\theta/2)}-i\cos{(\theta/2)}({\boldsymbol\sigma}\cdot {\hat n})}{2}\theta^{(1)}+\\
& +i\sin{(\theta/2)}({\boldsymbol\sigma}\cdot {\hat n}^{(1)})\label{eq:usu2e}
\end{split}
\end{equation}
Here, ${\hat n}^{(1)}$ and $\theta^{(1)}$ are the derivatives of ${\hat n}$ and $\theta$, respectively, with respect to the perturbation and precisely:
\begin{equation}
\resizebox{0.5\textwidth}{!}{
${\hat n}^{(1)}=\left(\dfrac{-m_x^2m_y^{(1)}+m_xm_ym_x^{(1)}}{\left(m_x^2+m_y^2\right)^{3/2}};\ \dfrac{m_y^2m_x^{(1)}+m_xm_ym_y^{(1)}}{\left(m_x^2+m_y^2\right)^{3/2}};\ 0 \right)$\label{eq:usu2f}
}
\end{equation}
\begin{equation}
\resizebox{0.5\textwidth}{!}{
$\theta^{(1)}=\dfrac{-\left(m_x^2+m_y^2\right)m_z^{(1)}+\left(m_xm_x^{(1)}+m_ym_y^{(1)}\right)m_z}{|m|^2\sqrt{m_x^2+m_y^2}}$ \label{eq:usu2g}
}
\end{equation}

Combining Eq.s (\ref{eq:usu2d}), (\ref{eq:usu2e}), (\ref{eq:usu2f}), (\ref{eq:usu2g}) and using the normalization conditions for the rotation axis components ($n_x^2+n_y^2=1$ and $n_xn_x^{(1)}+n_yn_y^{(1)}=0$), we recognize that:
\begin{equation}
U^{\dagger(0)}V_{xc}^{(1)}U^{(0)}=v_{xc}^{(1)}{\bf I}+B_{xc}^{(1)}\dfrac{\left({\boldsymbol\sigma}\cdot {\bf m}\right)}{m}\label{eq:usu2h}
\end{equation}
and
\begin{equation}
\begin{split}
& U^{\dagger(1)}V_{xc}^{(0)}U^{(0)}+U^{\dagger(0)}V_{xc}^{(0)}U^{(1)}=\\
& = B_{xc}^{(0)}\left[\left(\sin{(\theta)}\ n_y^{(1)}+\cos{(\theta)}\ n_y\theta^{(1)}\right)\sigma_x-\left(\sin{(\theta)} \ n_x^{(1)}+\right.\right.\\
& \left.\left.+\cos{(\theta)}\ n_x\theta^{(1)}\right)\sigma_y-\sin{(\theta)}\theta^{(1)}\sigma_z\right] \ .\label{eq:usu2i}
\end{split}
\end{equation}
The last two expressions constitute the final equations to treat the LSDA approximation of the XC functional. These equations, (\ref{eq:usu2h}) and (\ref{eq:usu2i}), have been implemented as an alternative to the method {\bf (i)} presented at the previous subsection, and which we identify as {\bf (i')}. 

\subsection{(ii) Explicit evaluation of the first order XC potential}
\label{sec:methodBXC}

Alternatively to the expansion of the rotation matrix, an explicit expression of the first order XC potential can be obtained by performing a Taylor expansion of the right hand side of Eq. (\ref{eq:pot0}). 
Introducing the LSDA definitions of the XC electrostatic potential $v_{xc}$ and the XC magnetic field magnitude $B_{xc}$,
\begin{equation}
\begin{split}
V_{xc} & = \frac{1}{2}\left( \frac{\partial E_{xc}}{\partial \rho_\uparrow}+\frac{\partial E_{xc}}{\partial \rho_\downarrow}\right)\\
B_{xc} & = \frac{1}{2}\left( \frac{\partial E_{xc}}{\partial \rho_\uparrow}-\frac{\partial E_{xc}}{\partial \rho_\downarrow}\right)
\end{split}
\label{eq:definitionBxc}
\end{equation}
one obtains the following expression for $V_{xc}^{(1)}$ relative to the global spin quantization axis (omitting again the subscript ``$xc$''):
\begin{equation}
\begin{split}
\hat{V}_{\alpha\beta}^{(1)} & =\frac{B^{(0)}_{xc}}{m}\left({\boldsymbol{\sigma}_{\alpha\beta}}-\frac{({\boldsymbol{\sigma}_{\alpha\beta}}\cdot\bf{m})\bf{m}}{m^{2}}\right)\cdot{\bf{m}}^{(1)}+ v^{(1)}_{xc}\ {\bf I}_{\alpha\beta}\  + \\
& + B^{(1)}_{xc} \; \frac{({\boldsymbol{\sigma}_{\alpha\beta}}\cdot{\bf{m}})}{m},
\end{split}
\label{eq:resultBXC}
\end{equation}
where ${\bf I}$ denotes the identity matrix. 
In the above expression, the first term corresponds to the change of the direction of the XC magnetic field due to rotation of magnetization induced by an external perturbation (e.g. atomic displacements). 
The second term describes the change of electric scalar potential induced by changes of $n$ and $m$, while the last term stems from the magnitude change of $B_{xc}$ due to the variation of the electronic density $n$ and the magnetic moment magnitude $m$. 
Hence, it can be readily recognized that the combination of the second and the third terms corresponds to a rotation of the collinear LSDA potential using the
$\hat{U}^{(0)}$ transformation matrix, while the first term, related to the change of magnetization direction, contains contributions involving the $\hat{U}^{(1)}$ transformation matrix introduced above. The second derivation is comparatively simple and uses more intuitive ingredients like the xc effective magnetic field.

Within the LSDA approximation, the spin diagonal $V^{(1)}_{xc}$ and $B^{(1)}_{xc}$ can be expressed as
\begin{equation}
\begin{split}
V^{(1)}_{xc} &= \frac{1}{2}\left( \frac{\partial^2 E_{xc}}{\partial \rho_\uparrow^2} \rho_\uparrow^{(1)} +
\frac{\partial^2 E_{xc}}{\partial \rho_\downarrow^2} \rho_\downarrow^{(1)} \right.\\
&+\left.
\frac{\partial^2
E_{xc}}{\partial \rho_\uparrow \partial \rho_\downarrow} (\rho_\uparrow^{(1)}+\rho_\downarrow^{(1)}) \right)\ ,\\
B^{(1)}_{xc} &= \frac{1}{2}\left( \frac{\partial^2 E_{xc}}{\partial \rho_\uparrow^2} \rho_\uparrow^{(1)} -  \frac{\partial^2 E_{xc}}{\partial \rho_{\downarrow}^2} \rho_\downarrow^{(1)}\right.\\
+ &\left.\frac{\partial^2 E_{xc}}{\partial \rho_\uparrow \partial \rho_\downarrow} (\rho_\downarrow^{(1)}-\rho_\uparrow^{(1)})\right).
\end{split}
\end{equation}
In these expressions, we use $\rho_\uparrow$ and $\rho_\downarrow$ which are defined in Eq.s (\ref{rhoupup}) and (\ref{rhodndn}). 
Although the expressions (\ref{eq:resROT}) and (\ref{eq:resultBXC}) should be equivalent, Eq. (\ref{eq:resROT}) is more generic since, in principle, it can be applied to any combination of $V_{xc}^{(0)}$ and $V_{xc}^{(1)}$. 

In the following, we will test and compare the two methods to verify that we obtain the same result, and see if one method is more efficient than the other.

\subsection{Implementation for atomic and electric field perturbations}\label{sec:tech}
We have implemented the aforementioned treatments of DFPT within non-collinear magnetism in the \textsc{abinit} code \cite{abinit2016} (available from version 8.8.3 onwards). This implementation is currently only available in the norm-conserving pseudopotential formalism; the generalization to the Projector Augmented-Wave (PAW) approach is in progress and should be straightforward: the spin rotation should be applied to the PAW on-site XC potential expressed in terms of spherical harmonics.
In the case of an atomic displacement perturbation, one has to take care of the so-called  ``frozen'' part of the second energy derivative (see Eq. (13) in Ref. \cite{Gonze1997}), which is not affected by the non-collinearity of the density.
Another delicate term is the so-called ``non-linear exchange-correlation core correction'' (NLCC)\cite{Bechstedt}, which is a correction to the XC energy to compensate the error caused by the frozen core approximation, and the separation between core and valence charge densities in the pseudopotential formalism, accounting for possible overlap between them. 
Even if this correction only concerns the electronic charge density $n$, it should be accounted for correctly because it is not invariant under spinor rotation transformations.
The treatment of electric field perturbation involves the derivative with respect to the k-point (ddk), which does not involve any frozen part or NLCC, such that its generalization into non-collinear magnetism is straightforward.
The implementation works also in the case where the spin-orbit coupling is included. 

The different methods of magnetization rotation can be set through the \verb|ixcrot| \textsc{abinit} input flag. 
The default value is method {\bf (i)} (\verb|ixcrot|=1).
The method {\bf (ii)} or {\bf (i')} can be used by setting \verb|ixcrot| to 2 or 3 respectively.

\section{Test cases}
\subsection{Accuracy and convergence of the implementation: Cr$_2$O$_3$}

\begin{table}[h!]
\tabcolsep=0.5em
\begin{tabularx}{\columnwidth}{c|c|c|c|c|c}
\toprule
\multicolumn{6}{c}{ }\\[-2px]
\multicolumn{6}{c}{\bf Frozen}\\[5px]
\toprule
& & & & &\\[-9px]
RM & {$\hat{\bf m}$} & E$^{(2)}$ (FD) & E$^{(2)}$ (DFPT) & $\Delta$V & \# SCI \\
\hline
\hline
& & & & &\\[-3px]
{\bf (i)} & \multirow{4}{*}{$\hat{x}$} & \multirow{4}{*}{26330.12639} & 26330.12833 & 9\ 10$^{-9}$ & 32\\[5px]
{\bf (i')} & & & 26330.12833 & 5\ 10$^{-9}$ & 40 \\[5px]
{\bf (ii)} & & & 26330.12833 & 5\ 10$^{-5}$ & 80 \\[5px]
\hline
& & & & &\\[-3px]
{\bf (i)} & \multirow{4}{*}{$\hat{y}$} & \multirow{4}{*}{26330.12639} & 26330.12833 & 1\ 10$^{-8}$ & 39\\[5px]
{\bf (i')} & & & 26330.12833 & 9\ 10$^{-9}$ & 39 \\[5px]
{\bf (ii)} & & & 26330.12833 & 1\ 10$^{-4}$ & 80 \\[5px]
\hline
& & & & &\\[-3px]
{\bf (i)} & \multirow{4}{*}{$\hat{z}$} & \multirow{4}{*}{26330.12640} & 26330.12833 & 7\ 10$^{-9}$ & 30\\[5px]
{\bf (i')} & & & 26330.12833 & 6\ 10$^{-9}$ & 30 \\[5px]
{\bf (ii)} & & & 26330.12833 & 8\ 10$^{-7}$ & 80 \\[5px]
\toprule
\multicolumn{6}{c}{ }\\[-2px]
\multicolumn{6}{c}{\bf Total}\\[5px]
\toprule
& & & & &\\[-9px]
RM & {$\hat{\bf m}$} & E$^{(2)}$ (FD) & E$^{(2)}$ (DFPT) & $\Delta$V & \# SCI \\
\hline
\hline
& & & & &\\[-3px]
{\bf (i)} & \multirow{4}{*}{$\hat{x}$} & \multirow{4}{*}{16.13695} & 16.13785 & 7\ 10$^{-9}$ & 36\\[5px]
{\bf (i')} & & & 16.14042 & 3\ 10$^{-9}$ & 39\\[5px]
{\bf (ii)} & & & 16.13785 & 3\ 10$^{-5}$ & 80\\[5px]
\hline
& & & & &\\[-3px]
{\bf (i)} & \multirow{4}{*}{$\hat{y}$} & \multirow{4}{*}{16.13693} & 16.13785 & 6\ 10$^{-9}$ & 39\\[5px]
{\bf (i')} & & & 16.14042 & 4\ 10$^{-9}$ & 39\\[5px]
{\bf (ii)} & & & 16.13785 & 7\ 10$^{-5}$ & 80\\[5px]
\hline
& & & & &\\[-3px]
{\bf (i)} & \multirow{4}{*}{$\hat{z}$} & \multirow{4}{*}{16.13694} & 16.13785 & 4\ 10$^{-9}$ & 32\\[5px]
{\bf (i')} & & & 16.14042 & 1\ 10$^{-8}$ & 40\\[5px]
{\bf (ii)} & & & 16.13785 & 3\ 10$^{-5}$ & 80\\[5px]
\hline
\end{tabularx}
\caption{Atomic displacement tests comparing FD to the three different transformation methods for DFPT: {\bf (i)}, {\bf (i')}, {\bf (ii)}. Values give the second derivatives of the energy with respect to reduced displacements (values in Ha). SCI is the number of the SCF cycles needed to converge with a tolerance $\Delta V$ (Ha). FD are performed using 5 discretization points for the Frozen part and 3 points for the Total one. Spin-orbit coupling is not included.}\label{tab:00A}
\end{table}

In this section we show how the different methods to treat the non-collinear DFPT perform for phonons with the LDA XC functional.
We start by performing tests on Cr$_2$O$_3$, which is a collinear antiferromagnetic insulator.
Cr$_2$O$_3$ is also known because it was the first proposed and measured magnetoelectric crystal \cite{dzyaloshinskii1959,astrov1960,astrov1961} and became a reference system to study the microscopic origin of bulk magnetoelectricity from DFT \cite{iniguez2008,bousquet2011, malashevich2012, tillack2016}, which requires non-collinear treatment of the magnetism (to go beyond the exchange-striction effect\cite{mostovoy2010}).

We perform our calculations with the \textsc{abinit} package and the Pseudodojo LDA norm-conserving pseudopotentials \cite{pseudodojo} (including spin-orbit coupling). Cr$_2$O$_3$ keeps its insulating state even without Hubbard correction for the d orbitals of the Cr atom.
To test our implementation of the DFPT within non-collinear magnetism, we first compare the second energy derivatives (E$^{(2)}=\frac{1}{2}\frac{\partial^2E}{\partial\lambda^2}$) obtained from DFPT and finite difference (FD) calculations for an atomic displacement perturbation in one direction (Cr atom along the $x$ direction displaced by four amplitudes - two positive and two negative: -2$\tau$, $-\tau$, $+\tau$ and $+2\tau$, with $\tau=0.003$ \AA{} - ; analogous results hold for the other two displacement directions that are not shown for simplicity). This 5-point FD scheme allows to extract the pure harmonic contribution to the energy variation and thus guarantees the validity of the comparison with the linear response calculations at the second order.
The results are shown in Tab. \ref{tab:00A} where we report the comparison of the frozen contribution alone and the total energy second derivatives for both DFPT (within its three  (i), (i') and (ii) methods) and FD. 
In Tab.~\ref{tab:00A} we also include the number of self-consistent iterations (SCI) needed to reach a given residual of the potential to stop the SCI. 
We also put a maximum number of iterations of 80 such that if the SCI did not reach the required precision the calculation will stop and we report the potential residual that the calculation could reach at the 80$^{\text{th}}$ step.
As can be seen the three approaches agree to within machine precision, and the agreement with FD is within 5-7 significant digits. 
We also observe that within 80 SCI steps, the method {\bf (ii)} does not reach the same precision as methods {\bf (i)} or {\bf (i')} do. 
Method {\bf (ii)} seems then to give slower converge. 
We also remark that method {\bf (i)} converges with a bit less number of SCI steps than method {\bf (i')}.
We note that this slower convergence of method {\bf (ii)} is observed even with changing the mixing or preconditioning parameters.
However, this characteristics may vary depending on the case study and method {\bf (ii)} could appear better in other crystals than Cr$_2$O$_3$.

In Tab.\ref{tab:Cr2O3} we report the calculated TO phonon frequencies of Cr$_2$O$_3$ for the collinear case and for the non-collinear (magnetic moments along the $z$ direction, the $x$ and $y$ directions give the same results and are not shown for clarity) case without spin-orbit coupling.
We can see that the method (i), (i') and (ii) give strictly the same phonon frequencies within less than 1 cm$^{-1}$. 
The comparison is also extremely good with the collinear case, which is what we expect for a non-frustrated collinear antiferroagnet as Cr$_2$O$_3$.
These very good agreement confirm the robustness of our implementation.

\begin{table}[ht!]
    \begin{tabular}{c|c|c|c|c}
    \hline
    \hline
    mode \# & Collinear  & non-coll. (i) & non-coll. (i') & non-coll. (ii) \\
    \hline
     1-2   & 234  & 234  & 234 & 234 \\
     3     & 236  & 236  & 236 & 236 \\
     4     & 246  & 246  & 246 & 246 \\
     5-6   & 279  & 280  & 280 & 280 \\
     7-8   & 300  & 300  & 300 & 300 \\
     9-10  & 357  & 357  & 357 & 357 \\
     11    & 361  & 361  & 361 & 361 \\
     12    & 382  & 382  & 382 & 382 \\
     13-14 & 414  & 414  & 414 & 414 \\
     15    & 437  & 437  & 437 & 437 \\
     16    & 447  & 447  & 447 & 447 \\
     17-18 & 464  & 465  & 465 & 465 \\
     19    & 466  & 466  & 466 & 466 \\
     20-21 & 473  & 473  & 473 & 473 \\
     22-23 & 522  & 522  & 522 & 522 \\
     24    & 537  & 537  & 537 & 537 \\
     25-26 & 545  & 545  & 545 & 545 \\
     27    & 567  & 567  & 567 & 567 \\
    \hline
    \end{tabular}
    \caption{Calculated phonon frequencies (cm$^{-1}$) of Cr$_2$O$_3$ for collinear case and non-collinear with magnetic moments along the $z$ directions (no SOC). We compare the results given by method (i), (i') and (ii), which all give the same result and are, as expected for collinear antiferromagnet, extremely close to the collinear case.}
    \label{tab:Cr2O3}
\end{table}

\subsection{Strong spin-orbit interaction case: RuCl$_3$}
$\alpha$-RuCl$_3$ is a layered material, with space group $P6_3/mcm$ and a honeycomb arrangement of the Ru atoms, where the 4$d$ electrons drive unusual electronic and magnetic behavior due to strong spin-orbit interaction \cite{plumb2014}.
The magnetic anisotropy of RuCl$_3$ is very strong, where the magnetic moments lie perpendicular to the stacking plane and where a magnetic field of about 14 T is necessary to flip the moments in-plane \cite{majumder2015}.
The magnetic ground state is known to be the so-called ZZ antiferromagnetism (ZZ-AFM) but for our test purpose we will simulate the crystal within its ferromagnetic phase.
As for Cr$_2$O$_3$, we performed our calculations with the \textsc{abinit} package and the Pseudodojo LDA norm-conserving pseudopotentials \cite{pseudodojo} including spin-orbit coupling.
The calculations were done with a grid of 8$\times$8$\times$8 k-points and a cutoff energy of 50 Ha for the plane wave expansion.

We performed internal atomic coordinate relaxations at fixed cell parameters ($a=$6.561 and $c=$5.709 \AA) for both collinear and non-collinear cases (with a residual on the forces of 2 10$^{-5}$ Ha/Bohr) before performing the DFPT calculations of the phonons (using method (i)).
For the non-collinear simulations, we treated both in-plane ($x$-direction) and out-of-plane ($z$-direction) alignment of the magnetic moments.
Since no Hubbard correction is done for the Ru-$d$ orbital correlations, we find that the ground state is metallic with or without the spin-orbit coupling (as reported in Ref. \cite{majumder2015}).
We obtain a magnetocrystalline anisotropy of about 780 $\mu$eV, which is consistent with previous calculations \cite{iyikanat2018} and the strong anisotropy reported experimentally.

\begin{table}[ht!]
    \begin{tabular}{c|c|c|c|c}
    \hline
    \hline
    mode \# & Label & Collinear  & non-collinear ($\hat{x}$) & non-collinear ($\hat{z}$) \\
    \hline
    1     & $\Gamma_2^+$ & 20$i$  & 19$i$  & 21$i$  \\
    2-3   & $\Gamma_5^+$ & 123    & 123     & 123  \\
    4     & $\Gamma_4^-$ & 146 & 146 & 147 \\
    5-6   & $\Gamma_5^-$ & 148 & 148 & 147  \\
    7     & $\Gamma_3^-$ & 169 & 166 & 158 \\
    8-9   & $\Gamma_6^-$ & 176 & 174 & 175  \\
    10-11 & $\Gamma_6^+$ & 249 & 249 & 248  \\
    12    & $\Gamma_4^+$ & 279 & 278 & 277 \\
    13-14 & $\Gamma_5^-$ & 283 & 283 & 282  \\
    15-16 & $\Gamma_5^+$ & 291 & 291 & 291  \\
    17    & $\Gamma_2^-$ & 298 & 298 & 296 \\
    18-19 & $\Gamma_6^-$ & 333 & 334 & 335 \\
    20    & $\Gamma_3^-$ & 355 & 347 & 351 \\
    21    & $\Gamma_1^+$ & 355 & 354 & 352 \\
    \hline
    \end{tabular}
    \caption{Calculated phonon frequencies (cm$^{-1}$) of $\alpha$-RuCl$_3$ in its FM phase for collinear case and non-collinear with magnetic moments along the $x$ and $z$ directions.}
    \label{tab:RuCl3}
\end{table}

We report the results in Tab.\ref{tab:RuCl3}.
We can observe that the differences between collinear and non-collinear cases are small. Frequency differences for different orientations are within $<2$ cm$^{-1}$, except the $\Gamma_3^-$ modes. A deviation of 9 cm$^{-1}$ is observed for mode number 7 when spins are along the $z$ directions and of 3 cm$^{-1}$ when spins are along the $x$-direction (when comparing to the collinear calculation). A deviation of 9 cm$^{-1}$ and 4 $^{-1}$ is observed for mode number 20 when spins are aligned along $x$ and $z$ respectively.
We also remark a small frequency shift of 1 (spins along $x$ direction) and 3 cm$^{-1}$ (spins along $z$ direction) of mode number 21 with label $\Gamma_1^+$.
The modes with $\Gamma_3^-$ label, which exhibit the strong non-collinear spin-phonon coupling, are the only two modes that involves motions of the Ru against each other along the $z$ direction. The associated Cl motions are along the in-plane direction only and are such that they come closer to Ru going away from their $xy$ plan and they go away from Ru going closer to their plan.
This spin-orbit shift of \emph{phonon} bands denotes an appreciable spin-electron-phonon coupling, which would require further studies to understand the exact microscopic origin.
We also observe a small unstable mode for all the cases, showing that the FM phase at the fixed cell parameters we used is structurally unstable (which is not surprising as it is not the ground state).

The results on RuCl$_3$ validate our implementation, in that performing  explicitly non-collinear DFPT calculations is important for systems where the non-collinearity is strongly coupled to the lattice properties of the crystal.

\section{Conclusion}
In this paper we have shown how to treat non-collinear magnetism within the DFPT.
The main problem is that most of the XC functionals, like LDA or GGA, are derived at the collinear level, thus missing the non-diagonal terms of the density containing the x and y components of the magnetization. This prevents to use them as such in DFPT, where perturbations can change the orientation of the magnetization.
We have derived several possibilities to treat this problem, which all are based on reducing the non-collinear derivatives to their collinear ones by aligning the z coordinate axis to the local magnetization direction (before or after perturbation). 
At each point the derivative of the XC potential can thus be performed in the usual collinear framework, the knack being to perform the back and forth transforms between the local collinear system and the global non-collinear one.
We propose two possibilities to perform these changes of coordinate system for the XC functional, either through a Taylor expansion of the rotation matrix between the two coordinate systems, or using an analytical expression of the XC functional (only done for the LDA here).
These different methods produce identical energy results in the test cases we explored, with method {\bf (ii)} being somewhat slower and harder to converge. In pathological cases it is possible that one method could be qualitatively better. 
We have done tests on the atomic displacement and electric field perturbations in Cr$_2$O$_3$ and shown that in RuCl$_3$ with strong spin-orbit coupling and magnetic anisotropy the phonons are affected by the non-collinear spin directions.

The immediate perspectives of our work are to extend this formalism to the Projector Augmented-Wave (PAW) approach \cite{audouze2006,audouze2008} and to GGA and hybrid XC functionals, and to combine it with the linear response of crystals to magnetic fields\cite{savrasov1998,cao2018}.

\begin{acknowledgments}
* SP and FR contributed equally to this work.
The authors would like to thank X. Gonze for the fruitful discussions.
EB MJV and FR acknowledge the FRS-FNRS for support through PDR projects MaRePeThe (GA 19528980) and ``Transport in novel VDW heterostructures'' (GA T.1077.15-1/7).
SP acknowledge the Marie Curie COFUND postdoctoral fellow at the University of Liege and with the support of the European Union.
MV, FR, SP and EB acknowledge the ARC AIMED project, the PRACE project TheDeNoMo and the CECI facilities funded by F.R.S-FNRS (Grant No. 2.5020.1) and Tier-1 supercomputer of the F\'ed\'eration Wallonie-Bruxelles funded by the Walloon Region (Grant No. 1117545).
MT was granted access to the french HPC resources of TGCC under the allocation 2018-AP010910358 attributed by GENCI (Grand Equipement National de Calcul Intensif).
\end{acknowledgments}
\appendix
\section{Building the density matrix}

\subsection{DFPT: first order}
In order to write the expression for the first order density matrix $\hat{\rho}^{(1)}$ we need to derive Eq. (\ref{eq:rho0}) with respect to a given perturbation $\lambda$:
\begin{eqnarray}
\hat{\rho}^{(1)}&=&
\frac{\partial\hat{\rho}^{(0)}}{\partial\lambda}=\dfrac{\partial}{\partial\lambda}\left(\ket{\Psi^{(0)}}\bra{\Psi^{(0)}}\right)\\
&=&\ket{\Psi^{(1)}}\bra{\Psi^{(0)}}+\ket{\Psi^{(0)}}\bra{\Psi^{(1)}}\ .
\end{eqnarray}
The generic element of this operator, projected on the $\alpha,\beta$ spin states, is:
\begin{eqnarray}
\rho^{(1)}_{\alpha\beta}&=&\braket{\alpha|\Psi^{(1)}}\braket{\Psi^{(0)}|\beta}+\braket{\alpha|\Psi^{(0)}}\braket{\Psi^{(1)}|\beta}=\\
&=&\psi_{\alpha}^{(1)}\psi_{\beta}^{(0)*}+\psi_{\alpha}^{(0)}\psi_{\beta}^{(1)*}\ .
\end{eqnarray}
At this point, we can write the $\hat{\rho}^{(1)}$ general expression, in terms of the wavefuctions analogously to the GS (see Eq. (\ref{eq:rhonew})):
\begin{eqnarray}
\hat{\rho}^{(1)}&=&\left( \begin{array}{cc}
\psi^{(1)}_1\psi_1^{(0)*} & \psi^{(1)}_1\psi_2^{(0)*} \\[10px]
\psi^{(1)}_2\psi_1^{(0)*} & \psi_2^{(1)}\psi_2^{(0)*} \end{array} \right)+\nonumber\\[1em]
&+&\left( \begin{array}{cc}
\psi^{(0)}_1\psi_1^{(1)*} & \psi^{(0)}_1\psi_2^{(1)*} \\[10px]
\psi^{(0)}_2\psi_1^{(1)*} & \psi_2^{(0)}\psi_2^{(1)*} \end{array} \right)=\\[10px]
&=&\left( \begin{array}{cc}
\psi^{(1)*}_1\psi_1^{(0)}+ \psi^{(0)*}_1\psi_1^{(1)} & \psi^{(1)}_1\psi_2^{(0)*}+\psi^{(0)}_1\psi_2^{(1)*} \\[10px]
\psi^{(1)}_2\psi_1^{(0)*}+\psi^{(0)}_2\psi_1^{(1)*} & \psi_2^{(1)*}\psi_2^{(0)} +\psi_2^{(0)*}\psi_2^{(1)}\end{array} \right)=\nonumber\\[10px]
&=&\dfrac{1}{2}\left( \begin{array}{cc}
\rho^{(1)} + m^{(1)}_z & m^{(1)}_x-i\ m^{(1)}_y \\[10px]
m^{(1)}_x + i\ m^{(1)}_y & \rho^{(1)}-m^{(1)}_z \end{array} \right)\ .
\end{eqnarray}
The last equation has been obtained as a derivative of the Eq. (\ref{eq:ropaul}) and, interestingly, we note that the shape of the density matrix does not change with respect
a given order of perturbation. 
Following the analogous procedure used for the GS case (Eq. (\ref{eq:density0}) and following), we obtain the expression of the charge and magnetization density components in terms of spinors:
\begin{eqnarray}
\rho^{(1)}&=&\left[\psi^{(1)*}_1\psi_1^{(0)}+ \psi^{(0)*}_1\psi_1^{(1)}+\right.\nonumber\\
&&   +\left.\psi_2^{(1)*}\psi_2^{(0)} +\psi_2^{(0)*}\psi_2^{(1)}\right]\label{eq:densityel0}\\
m^{(1)}_x&=&\left[\psi^{(1)*}_1\psi_2^{(0)}+\psi^{(0)*}_1\psi_2^{(1)}+\right. \nonumber\\
&&   +\left.\psi^{(1)*}_2\psi_1^{(0)}+\psi^{(0)*}_2\psi_1^{(1)}\right]\\
m^{(1)}_y&=&i\left[\psi^{(1)}_1\psi_2^{(0)*}+\psi^{(0)}_1\psi_2^{(1)*}+\right. \nonumber\\
&&   -\left.\left(\psi^{(1)}_2\psi_1^{(0)*}+\psi^{(0)}_2\psi_1^{(1)*}\right)\right]\\
m^{(1)}_z&=&\left[\psi^{(1)*}_1\psi_1^{(0)}+ \psi^{(0)*}_1\psi_1^{(1)}+\right. \nonumber\\
&&   -\left.\left(\psi_2^{(1)*}\psi_2^{(0)} +\psi_2^{(0)*}\psi_2^{(1)}\right)\right]\ .
\end{eqnarray}
Separating the real and imaginary parts of the wavefunctions to rewrite the last equations we remark that all the quantities are real, as we expected:
\begin{eqnarray}\label{eq:densityel}
\rho^{(1)}&=&2\left[\psi_1^{(0)'}\psi^{(1)'}_1+\psi_1^{(0)''}\psi^{(1)''}_1\right.+\nonumber\\
&&   +\left.\psi_2^{(0)'}\psi^{(1)'}_2+\psi_2^{(0)''}\psi^{(1)''}_2\right]\\
m_z^{(1)}&=&2\left[\psi_1^{(0)'}\psi^{(1)'}_1+\psi_1^{(0)''}\psi^{(1)''}_1\right.+\nonumber\\
&&   -\left.\psi_2^{(0)'}\psi^{(1)'}_2+\psi_2^{(0)''}\psi^{(1)''}_2\right]\\
m_x^{(1)}&=&2\left[\psi_1^{(1)'}\psi_2^{(0)'}+\psi_1^{(0)'}\psi_2^{(1)'}+\right. \nonumber \\
&&   +\left.\psi_1^{(0)''}\psi_2^{(1)''}+\psi_2^{(0)''}\psi_1^{(1)''}\right]\\
m_y^{(1)}&=&2\left[\psi_1^{(1)'}\psi_2^{(0)''}-\psi_1^{(1)''}\psi_2^{(0)'}+\right. \nonumber \\
&&   +\left.\psi_1^{(0)'}\psi_2^{(1)''}-\psi_1^{(0)''}\psi_2^{(1)'}\right]\ .
\end{eqnarray}

\section{Local potential integral}

In addition to the perturbed density matrix elements, that have to be built for a non-collinear magnetic system, we need to specify the terms involved in this complicated formalism, as well as the so-called ``local potential''. This term (see Eq. (91) in \cite{Gonze1995b}), for a given band, is defined as follows:
\begin{eqnarray}
\braket{\psi^{(1)*}|(H-\varepsilon)^{(0)}|\psi^{(1)}}=\braket{\psi^{(1)*}|v^{(0)}|\psi^{(1)}}=\\
=\int \psi({\bf r})^{(1)*}v({\bf r})^{(0)}\psi({\bf r})^{(1)}d({\bf r}) .
\end{eqnarray}
The terms to be integrated then can be explicitly written performing the matrix product between spinors and the local potential matrix:
\begin{eqnarray}
=\int &d{\bf r}&\left( \begin{array}{cc}
\psi^{(1)}_1 & \psi^{(1)}_2 \end{array}\right)^*
\left( \begin{array}{cc}
v_{11} & v_{12} \\[10px]
v_{21} & v_{22} \end{array} \right)
\left( \begin{array}{c}
\psi^{(1)}_1 \\[10px]
\psi^{(1)}_2 \end{array}\right)=\\
=\int &d{\bf r}&\left[v_{11}\left|\psi_1\right|^2+v_{22}\left|\psi_2\right|^2+2v'_{12}\psi'_1\psi'_2\right. +\nonumber\\
&+&\left.2v'_{12}\psi''_1\psi''_2-2v''_{12}\psi'_1\psi''_2+2v''_{12}\psi''_1\psi'_2\right]
\end{eqnarray}

\end{document}